\documentclass[twocolumn,preprintnumbers,amsmath,amssymb]{revtex4}
\usepackage{graphicx}
\usepackage{epsfig}
\usepackage{amsmath}
\usepackage{amsfonts}
\usepackage{amssymb}
\usepackage{textcomp}
\usepackage{dcolumn}

\begin{document}

\title{Copenhagen Interpretation of Quantum Theory and the Measurement Problem}

\author{Martin Kober}
\email{kober@th.physik.uni-frankfurt.de}

\affiliation{Institut f\"ur Theoretische Physik, Johann Wolfgang Goethe-Universit\"at, Max-von-Laue-Str.~1, 
60438 Frankfurt am Main, Germany}


\begin{abstract}
The Copenhagen interpretation of quantum theory is investigated from a philosophical point of view. It is justified the opinion that the philosophical attitude the Copenhagen interpretation is based on is in principle inevitable for a real comprehension of quantum theory. This attitude is mainly related to epistemological arguments. However, the measurement problem often seems not to be treated clearly enough within the interpretation. By referring to the property of the necessity to use macroscopic measurement instruments obeying classical concepts it is made the attempt to solve the measurement problem. According to this consideration the indeterministic character of quantum theory seems to have its origin in a lack of knowledge and thus it appears in a similar but more principle way than in statistical mechanics. It is emphasized the ontological character of the uncertainty relation and the related non locality of quantum theory suggesting that the existence of a position space is not as fundamental as the assumptions of general quantum theory.
\end{abstract}

\maketitle

\section{Introduction}

Since the nineteen-twenties quantum theory can be considered as the most fundamental description of nature being accessible so far, at least in the microscopic range. All matter and interaction fields of the standard model of particle physics are described within the framework of relativistic quantum field theories. Even general relativity is assumed to obey a quantum theoretical description on a fundamental level. However, the interpretation of quantum theory is still ambiguous. The Copenhagen interpretation can probably be regarded as the standard interpretation of quantum theory. But there arises the question if this interpretation is really understood and if the philosophical attitude of Werner Heisenberg and Niels Bohr is received accurately.
In this paper there is hold the opinion that the Copenhagen interpretation is inevitable at least as a necessary condition to obtain a real comprehension of quantum theory. To justify this view it is important to emphasize the decisive philosophical attitude it is based on. There will be given some fundamental philosophical reasons suggesting that it is mandatory concerning all further considerations as the treatment of the measurement problem. 
The core of the Copenhagen interpretation is founded in Bohr's statement that all phenomena which can appear in human experience, no matter if they are everyday phenomena or if they refer to the atomic or subatomic range, have to be described by classical concepts. Concerning an adequate comprehension of this constitutive assertion of Bohr that classical concepts are indispensable for a description of experiments it will be important to distinguish very carefully between ontological and epistemological arguments. I think that subjectivistic misinterpretations of Heisenberg's and Bohr's attitude have their origin in a missing accuracy concerning the distinction between this two sorts of arguments.\\
Independent of this, the measurement problem of quantum theory seems to be the decisive aspect concerning the question of consistence of quantum theory. It is related to the two different ways a state can change with time according to the postulates of quantum theory, with respect to the time dependent Schroedinger equation, if no measurement is performed, and spontaneously, if a measurement is performed. Together with the composition rule these two ways a state can change seem to imply an inherent contradiction of quantum theory. Therefore it seems to be necessary to explore the fundamental postulates of the abstract Hilbert space formulation of quantum theory concerning their relation to each other on the one hand and their status of fundamentality on the other hand.\\
As an attempt to circumvent this inherent ambiguity of quantum theory it will be suggested that the element of indeterminacy within quantum theory does not represent a fundamental property of nature in a strict ontological sense but has its origin in the necessity to use macroscopic objects as measurement instruments on which classical concepts can be applied. Here the decisive argument is that it is impossible to know all degrees of freedom of macroscopic objects being used as measurement instruments but that these degrees of freedom have an important influence on the way the state changes during the interaction between the measured object and the measurement instrument.
In this sense the element of indeterminacy in quantum theory is interpreted similar to the indeterminacy within statistical mechanics but in a much more principle way, since in statistical mechanics it is in principle always thinkable to omit this statistical element whereas in quantum theory it is mandatory because of the necessity to use classical concepts for a description of an experiment. This necessity to use classical concepts has its origin in the special constitution of human perception which is based on these concepts although they do not correspond to nature, at least not exactly, what can be considered as the reason for the epistemological origin of the necessity to use classical concepts.\\
Besides, it will be investigated the closely related question of the ontological status of the uncertainty relation and the quantum states and it will be given an interpretation of the non locality of quantum theory. Because of the fact that the dynamics of quantum theory takes place within an abstract Hilbert space which can be represented by referring to the position space or the momentum space but does not presuppose it in its most abstract formulation, the assumption seems to be plausible that the existence of a position space is a derived quantity which means that the position space we live in is the consequence of a certain representation of the dynamics of quantum theory. The effects where non locality appears explicitly, in the double slit experiment performed with single particles for example, or the correlation of spin states of particles with a large distance and the corresponding Einstein Podolsky Rosen paradox \cite{Einstein:1935}, can be considered as natural according to this attitude which is therefore strongly supported by them. The local description of classical physics seems to represent only an approximation where the representation in position space works very well. Concerning this consideration it is decisive that the validity of the uncertainty relation is independent of the indeterminacy in the description of quantum theory. The indeterminacy within the description of nature by quantum theory has its origin in accordance to the above consideration in the necessity to use macroscopic measurement instruments perturbing the state of a particle in way leading to a state accurately localized. The localization is necessary because all measurements are based on measurements of position from which other quantities are derived. But this does not imply that the uncertainty relation itself implies an indeterministic description in a fundamental ontological sense, because the evolution of the quantum states which have not to be localized very sharply are governed by the Schroedinger equation yielding a completely deterministic description. Because of these reasons there is suggested a strict ontological interpretation of a quantum theoretical state represented as a wave function implying that the measured classical variables are only classical approximations.

\section{The Copenhagen Interpretation of quantum theory}

In general it is hold the opinion that the Copenhagen interpretation is accepted as the established interpretation of quantum theory. But the Copenhagen interpretation seems to be ambiguous by itself and thus there might exist the necessity to give an interpretation of itself to obtain clearness about its real philosophical attitude.
With respect to this it is decisive to be aware of the fact that the philosophical position of the Copenhagen interpretation is essentially based on epistemological arguments. Often it seems to be interpreted in a way to much subjectivistic and positivistic. Such a kind of misapprehension seems to arise from a confusion between epistemological and ontological arguments. 
These two sorts of arguments have to be distinguished very carefully, if one deals with the interpretation of quantum theory. It will be made the attempt to give a justification of the attitude of Heisenberg and Bohr by relating it to the epistemology of Immanuel Kant. According to Carl Friedrich von Weizsaecker there exists a very close relation between the thinking of Kant and Bohr \cite{Weizsaecker:1971}. Let us begin with a citation of Heisenberg. Heisenberg's own description of the Copenhagen interpretation begins with the following sentences \cite{Heisenberg:1958}:

\begin{quote}
"`The Copenhagen interpretation of quantum theory begins with a paradox. Every physical experiment, no matter if it
refers to phenomena of daily live or to atomic physics, has to be described by the concepts of classical physics. The
concepts of classical physics represent the language in which we describe the configuration of our experiments and
determine the results. We cannot replace them by other concepts. Nevertheless the applicability of these concepts is
limited due to the uncertainty relations."'

Werner Heisenberg, Physik und Philosophie ("`Physics and Philosophy"'), 1958 \footnote{All the citations of this paper are translations from German texts being performed by myself.}.
\end{quote}
Here we have the decisive statement of the Copenhagen interpretation referring to the necessity to use classical concepts.
But it is equally important that the applicability of these concepts is limited according to the uncertainty relation representing the decisive property of quantum theory. Because of this tension the distinction between epistemology and ontology becomes very important. With respect to this consideration it is decisive to emphasize the mentioned fact that the uncertainty relation has an ontological status independent of human knowledge within quantum theory. The positivistic assertion that an exact position and an exact momentum cannot exist at the same time because it is impossible to measure them does not belong to the Copenhagen interpretation. Heisenberg has found the uncertainty relation by remembering a conversation with Albert Einstein where Einstein has emphasized the opposite point of view that a theory is necessary to determine which quantities can be measured. The argumentation of Heisenberg was that there do not exist states where an exact position and an exact momentum is defined at the same time according to the postulates of quantum theory and therefore it has to be impossible to measure them. His thought experiment with the photon showing the impossibility to measure both quantities exactly just served to reply to the objection that it seemed to be possible to measure them \cite{Heisenberg:1927,Heisenberg:1969,Weizsaecker:1999}.\\
The uncertainty can further not be interpreted as an incompleteness of quantum theory because there are many experiments like the double slit experiment where the assumption of an exact trajectory of a particle would cause contradictions. The interference pattern of the double slit experiment with single photons, for example, can only be interpreted, if it is not determined through which slit the trajectory of the photon runs. Especially the phase relations being not measurable directly are constitutive to understand this phenomenon. This means that nature does not obey a description by classical concepts like a trajectory of a particle. 
(A theory of hidden variables, discussed in \cite{Bell:1964fg} for example, cannot be excluded, but seems to be implausible,
because in this case the non classical phenomena in the atomic and subatomic range have to be derived from an underlying description whereas these phenomena arise naturally from usual quantum theory.)\\ 
Nevertheless, the Copenhagen interpretation asserts that these concepts are indispensable for a description of experiments. Why is it impossible to replace them by other concepts ? The answer to this question cannot be found in nature itself. It can only be found in the way of human perception.\\
The reason why we have to describe nature by using classical concepts has its origin in the way we have experience and thus in our relation to nature and not in nature itself. According to the philosophy of Kant space and time are basic constituents of human experience \cite{Kant:1781,Kant:1783}. They are preconditions to have experience (at least human experience) and are therefore constitutive for having experience, because it is not even possible to imagine a perception which does not contain space and time as basic shapes of experience. From this decisive epistemological insight he concluded that they are given a priori before any experience and thus represent basic properties of human perception being indispensable for human mind to obtain access to the real world. Causality was interpreted in a similar way by Kant.\\
The concepts of classical mechanics are in accordance with this structures of human perception given a priori. They correspond to the concepts of our mind. Quantum theory describes reality in a way being much more general. The uncertainty relation limits the validity of the classical concepts in the real world. But nevertheless they remain constitutive for human experience. Thus Bohr's statement concerning the necessity of classical concepts is closely related to the attitude of Kant. Kant could not anticipate the development of modern physics which led Bohr to the insight that there exist phenomena being not in accordance with the structures of human perception but which are nevertheless reflected within this structures, if human beings describe
experiments. Bohr's concept of complementarity arises from this relation between the limitation of the classical concepts and the inevitability to use them in experience. Classical concepts are not suitable to describe the phenomena. But by using several concepts which contradict each other within the realm of our classical thinking we can obtain an understanding of the reality behind these concepts which are only able to yield an idea of this greater reality which is represented inadequately by using them. An elementary particle like the electron for example is nothing which can be object of human experience with respect to its real nature. It is just possible to percept its representations within the structures of human experience and human thinking where electrons can appear as particles or waves depending on the situation of observance but not with respect to their real nature which is not recognizable for human beings. Concerning these considerations a citation of von Weizsaecker seems to be very illuminative \cite{Weizsaecker:1971}:

\begin{quote}
"`The parallelism of the two attitudes is all the more remarkable, since Bohr never seemed to have read Kant extensively. As distinguished from Kant, Bohr learned from modern atomic physics that there exists science beyond the scope where processes can be described reasonably through properties of objects being independent of the situation of the observer. This is expressed by his idea of complementarity."'

Carl Friedrich von Weizsaecker, Die Einheit der Natur ("`The Unity of Nature"'), 1971. 
\end{quote}
By incorporating the idea of biological evolution to the philosophy of human knowledge of Kant classical concepts, especially space and time, appear as real entities again, because the structures of human knowledge have been adapted to the structures of the real world to enable surviving \cite{Lorenz:1970, Vollmer:1975}. But this idea of the evolutionary philosophy of human knowledge just implies that the structures of human mind correspond to the real world in a way being accurately enough to enable surviving but not to yield objective truth in an absolute sense. Thus the description of physics in terms of classical concepts, as particles within a position space for example, is true in the sense that it represents an approximation of the structure of the real world but it has not to be true in an absolute sense as modern physics, with respect to quantum theory as well as to special and general relativity, has shown. This is the reason why we have to use classical concepts on the one hand, at least concerning the description of experiments, and why they are not in accordance with the structures of the real world on the other hand.\\
The philosophical attitude of the Copenhagen interpretation arises from this duality between the real world and our mind. If the concepts of our thinking correspond to the explored phenomena very closely like this is the case in classical physics, we can neglect the influence of our mind to our perception. But in quantum theory these concepts do not work any more and therefore we have to refer to the relation of our mind to nature. This does not mean that the behaviour of nature depends on the way we describe it or that our consciousness has an influence to the result of a measurement. This would be an enormous misinterpretation of the Copenhagen interpretation. But it means that if we describe nature, we have do it as human beings and therefore our description of nature(not nature itself) is also determined by and related to the inherent structures
of human perception.

\section{The postulates of quantum theory}

Let us give a short review about the postulates of general quantum theory formulated in terms of the abstract Hilbert space representation of Paul Dirac and John von Neumann \cite{Dirac:1958, Neumann:1932}. This will be helpful to formulate the core of the measurement problem explicitly and to question the meaning of the different postulates concerning their status of fundamentality. In the sequel of the next section it will be referred to this classification of the postulates of quantum theory.\\
\\
\noindent
{\bf postulate 1:}
\newline
A physical system is described by a vector $| \psi \rangle$ within an abstract complex Hilbert space $\mathcal{H}$ being a vector space which is
endowed with an inner product $\langle \cdot|\cdot \rangle$.
\newline\newline
{\bf postulate 2:}
\newline
An observable quantity is described by a linear Hermitian operator A within the Hilbert space $\mathcal{H}$. This means that
the eigenvalues of an operator representing such an observable quantity are the possible values one can
get by performing a measurement. Because the Operator is Hermitian, meaning that $A=A^\dagger$, the eigenvalues are real.
\newline\newline
{\bf postulate 3:}
\newline
There is specified a special operator H belonging to a certain system and determining the time evolution of a state. This is called Hamiltonian. If there takes place no measurement (interaction with a macroscopic system), then the state of the system evolves with time according to the Schroedinger equation $i \partial_t | \psi \rangle = H | \psi \rangle$ which expresses that the application of the general time translation operator to a state is equal to the application of the Hamiltonian being a function of the observable quantities determining the state or the system. This dynamical evolution can equivalently be formulated within the Heisenberg picture where the observable quantities depend on time according to the commutator with the Hamiltonian $\frac{d}{dt}A=\frac{1}{i\hbar}[A,H]_-$.
\newline\newline
{\bf postulate 4:}
\newline
If there is performed a measurement of a certain quantity, then the vector changes into an eigenvector of the corresponding operator and the measured value is the eigenvalue to the corresponding eigenstate. It is not determined to which eigenstate the vector will convert. But the probability $p(| \psi \rangle \rightarrow |a \rangle)$ to convert to a special eigenstate corresponds the square of the value of the inner product(defining the Hilbert space structure of postulate 1) between the state immediately before the measurement and the corresponding eigenstate 
$p(| \psi \rangle \rightarrow |a \rangle)=|\langle a | \psi \rangle|^2$.
\newline\newline
{\bf postulate 5:}
\newline
If there are two systems described by two vectors, $| \varphi \rangle$ and $| \chi \rangle$, within two different Hilbert spaces, then the two systems can be combined and the complete system is described by the tensor product of these states $|\varphi \rangle \otimes | \chi \rangle $.
\newline

Regarding the postulates of quantum theory it is very important to realize that quantum theory in this abstract setting does not presuppose a position or a momentum space. The fact that time seems to be more fundamental than space does not contradict special relativity, because special relativity asserts only that time and space are combined with each other in a different way than in classical mechanics but it does not abolish the difference between space and time. If one refers to a special inertial system, there is specified a time direction in a distinguished way and there still exists a causal structure being independent of the inertial system. However, the question of time is very difficult and will not be treated in this consideration. It shall just be mentioned that it seems not to be possible to omit the concept of a real time. All laws of physics presuppose the concept of time which is the most fundamental concept of human thinking. Even the existence of consciousness presupposes time but no space \cite{Kant:1781,Kant:1783}. Since it is not possible to justify the second law of thermodynamics without a reference to time, because the concept of probability presupposes the concept of time, it seems not to be possible to derive the existence of a time direction distinguished from the space directions from thermodynamics for example \cite{Weizsaecker:1985,Weizsaecker:1992}.

\section{The core of the measurement problem as the core of the consistence problem of quantum theory}

The decisive conceptual problem of quantum theory is related to the two different ways a quantum state can change according to the postulates of quantum theory. If there is performed no measurement (there takes place no interaction with a macroscopic instrument), a quantum state evolves with respect to the time dependent Schroedinger equation. This is expressed by postulate 3. In the Heisenberg picture the observable quantities evolve instead of the state according to the commutator with the Hamiltonian.
The description of the time evolution is completely deterministic. But if there takes place a measurement, the state is changed instantaneously and this takes place in an indeterministic way and the state only determines the probability of a conversion into an eigenstate of the measured variable what is expressed by postulate 4.
This two ways a state can change according to quantum theory contained within the postulates 3 and 4 in the above classification do not seem to be commensurable, if one additionally assumes that postulate 5, the composition rule, is correct. This has its origin in the following argumentation:\\
If one considers a system $S_1$ which state is represented by a vector $| \psi(t) \rangle$, then if one performs a measurement of the quantity described by an operator A at the time t, the state is converted immediately to an eigenstate $| a \rangle$ with a probability given by the squared inner product between the state immediately before the interaction with the measurement instrument $M$ and the eigenstate $|\langle a | \psi(t) \rangle|^2$. This description is not deterministic. Random gets a meaning in a principle sense. But with respect to this fact one has to realize that measurement means interaction with a (macroscopic) system. And if quantum theory is assumed to yield a fundamental description of nature, or at least a more fundamental description than any classical theory like classical mechanics referring to macroscopic systems, $M$ has to be ruled by quantum theory also. It may be that it is enough to describe it with a classical theory, but such a theory itself is considered as an approximation to quantum theory, if one assumes that quantum theory is fundamental in the mentioned sense.
Thus M itself can be treated as a quantum theoretical system $S_M$ and therefore its state can also be described as a vector 
$| \Psi(t) \rangle$ within a Hilbert space. And this implies according to postulate 5 that one can compose the system of the observed object $S_1$ considered originally being described by the state $| \psi(t) \rangle$ and the system of the measurement instrument $S_M$ represented by the state $| \Psi(t) \rangle$ to one system $S_C$ described by the tensor product of these two states $| \Psi(t) \rangle \otimes | \psi(t) \rangle$. But the dynamics of this composed system $S_C$ is described by a Hamiltonian determined by the two systems and therefore the evolution of this state is described deterministic as long as there is performed no measurement with respect to a new measurement system referring to the composed system. The Hamiltonian of the composed system $H_C$ can be written as the sum of the free Hamiltonian of the measured system $H_\psi$, the free Hamiltonian of the measurement instrument $H_\Psi$ and an interaction Hamiltonian $H_I$

\begin{equation}
H_C=H_\psi+H_\Psi+H_I
\end{equation}
and determines the dynamics of the composed system according to the corresponding Schroedinger equation

\begin{equation}
i \partial_t \left(|\Psi(t) \rangle \otimes |\psi(t) \rangle \right)=H_C \left( | \Psi(t) \rangle \otimes | \psi(t) \rangle \right).
\end{equation}
Since the measurement process represents an interaction between $S_1$ and $S_M$, this implies a dynamical and thus deterministic description of the measurement process. 
This means that in the first case we seem to have an indeterministic description of the system $S_C$ consisting of $S_1$ and $S_M$, since the state of the measured object changes spontaneously, and in the second case we seem to have a deterministic description of the same situation where $S_1$ and $M$ are considered as one big system $S_C$ with respect to a new measurement instrument. And this implies that quantum theory seems to contradict itself with respect to this ambiguity. The above argumentation can also be found in \cite{Rovelli:1995fv,Rovelli:2004}.\\
Thus the conceptual problem within quantum theory is not that its description of nature is non causal or that it is non local.
This aspects contradict the structures of human knowledge but this is nothing which makes a theory seeming to be implausible. I would almost assert that it is the other way round. The fact that the concepts of a theory are not in accordance with our imagination is not only no objection, it is even to be expected that the behaviour of nature on a very fundamental level deviates from the structures of human thinking. If one deals with the philosophy of human knowledge by incorporation of the biological evolution according to to the epistemological consideration given within the above description of the Copenhagen interpretation, one should not be astonished that the structures of the real world are much different than the structures of human thinking and having experience. This is because of the fact that they have developed according to the advantage they yielded human beings in surviving and not according to the aim of obtaining objective knowledge about the world.
Thus the real conceptual problem of quantum theory is that it seems to contradict itself with respect to the two different ways a state can change according to postulate 3 and postulate 4. Concerning this point one deals with an inherent contradiction of the theory, not a contradiction to our imagination but between the postulates of the theory itself. Thus it appears as one central task concerning a deeper comprehension of quantum theory to reconcile postulate 3 and 4 with respect to postulate 5. This means at least that these postulates have to be reinterpreted with respect to their statement about the fundamental behaviour of nature.

\section{Attempt of a reconciliation by referring to the necessity to use macroscopic measurement instruments to observe a quantum system}

If one accepts the contradiction at this stage of comprehension between postulate 3 on the on hand and postulate 4 under incorporation of postulate 5 on the other hand which becomes manifest within the above consideration, there arise several possibilities for a reconciliation. One possibility could be that one of the two postulates referring to the changing of states in quantum theory is wrong. Another possibility could be that postulate 5 is wrong. But this assumption that one of these postulates is simply wrong is not very plausible because the postulates are confirmed empirically very well. Besides, the several postulates of quantum theory belong together, because they build one closed framework for the description of nature. If one of the postulates were not correct, then one should expect that the whole framework of quantum theory would be relativized by a completely new theory being more fundamental and containing it as an approximation, but it is not very plausible to assume that quantum theory being very successful concerning the area of experience where it is assumed to be valid should be changed by itself.
Thus there arises the task to obtain a better comprehension of the meaning of the postulates concerning their relation to each other suggesting another possibility which consists in the assumption that one of the postulates has to be reinterpreted in a certain way. Carlo Rovelli has suggested the so called relational interpretation of quantum theory \cite{Rovelli:1995fv,Rovelli:2004,Smerlak:2006gi}. According to this interpretation a state is only defined with respect to a certain observer. The state of the system with respect to an observer who performs a measurement of a system $S_1$ with a measurement instrument $M_1$ has to be distinguished from the state with respect to another observer who performs a measurement of the from $S_1$ and $M_1$ composed system with another measurement instrument $M_2$. Thus the measurement problem is solved in a way seeming to be similar to the attitude special relativity is based on. In my opinion this approach represents a very interesting way to reconcile the above postulates.\\ 
However, there exists another possibility to cope with the problem. This possibility is preferred in this paper, because it is more conservative. This means that it needs no additional precondition with respect to the postulates of quantum theory.
It consists in the assumption that one of the two postulates represents not a fundamental description of nature but serves as an effective description with respect to the other postulate from which it can in principle be derived. In this paper this possibility is advocated and accordingly it is assumed that postulate 4 is a consequence of postulate 3. And this attitude seems to be quite plausible, if one remembers the central postulate of the Copenhagen interpretation of quantum theory: Experiments have to be described by classical concepts implying that a measurement instrument has to be a macroscopic system. Such a system obeys quantum theory but it can effectively be described in a classical way. To enable a classical treatment in an approximative way, the measurement instrument has to consist of a very large number of particles and this number has to be as large that it becomes impossible to know its state exactly consisting of the states of all its constituents and their entanglement.
But if the exact state of the measurement instrument cannot be known exactly, it is also impossible to describe the interaction between the measured object and the measurement instrument in an exact way, because the reaction of the measured system (describing perhaps only one particle) will be very sensitive to very small differences of the exact states of the particles the measurement instrument consists of and the quantum system is entangled with during the measurement process. 
This is in accordance with the phenomenon of decoherence appearing as a consequence of correlations of quantum systems with the environment which cause an annihilation of phase relations and thus interference effects and lead to a transition to classical properties. Considerations of decoherence concerning the interpretation of quantum theory can be found in \cite{Schlosshauer:2003zy} and in \cite{Schlosshauer:2008} there is especially regarded the relation of decoherence to Bohr's
decisive statement of the inevitability of classical concepts.
Thus there seems to be offered the possibility that postulate 4 could indeed be derived from postulate 3. In a fundamental description the measurement process would therefore be deterministic, but there would be only accessible an effective description because of the fact that the measurement instrument has to be described as a classical system to be accessible for human perception.
If this assumption that postulate 4 can in principle be derived from postulate 3 were correct, then quantum theory would be deterministic in a rigorous ontological sense. The element of indeterminacy in quantum theory would be similar to the role it plays in statistical mechanics, where random arises from the lacking knowledge of the constituents of the system. However, the situation in quantum theory would be a little bit different nevertheless, because in quantum theory it is a direct consequence of the way human beings have experience. To describe the status of indeterminacy corresponding to this interpretation of quantum theory we should therefore first classify different stages on which an element of indeterminacy can appear in a physical theory. Concerning this we will distinguish between the following meanings of indeterminacy within a physical theory:

\noindent
{\bf 1 no indeterminacy (classical mechanics, classical electrodynamics for example):}
\newline
In such a kind of theory there exists no element of indeterminacy. The dynamical evolution of all constituents necessary to yield a complete description
of a physical system described by the theory is determined.
\newline\newline
{\bf 2 indeterminacy in an epistemological sense having practical reasons (statistical mechanics for example):}
\newline
In such a kind of theory the appearing indeterminacy has to do with a lack of knowledge by practical reasons. It is not possible to know all parameters determining the dynamics of a system because of its complexity. But in principle it would be possible to measure all the constituents determining the dynamical evolution and a deterministic description is thinkable in principle.
\newline\newline
{\bf 3 indeterminacy in an epistemological sense having principle reasons (the interpretation of this paper to reconcile postulate 3 and 4 for example):}
\newline
In such a kind of theory the indeterminacy also arises from a lack of knowledge. But in contrast to the above case it is in principle impossible to get the information being necessary for a deterministic description. The nature of human perception excludes a deterministic description in principle. (One has to use a macroscopic system to measure the state of a particle for example.)
\newline\newline
{\bf 4 indeterminacy in an ontological sense (quantum theory with postulate 4 assumed to be fundamental for example):}
\newline
There exists an element of real indeterminacy within such a theory. This has nothing to to with our way to describe the world and the impossibility to obtain exact knowledge of all constituents describing the system but has an ontological status.
\newline\newline
As already indicated within the brackets the interpretation of the quantum theoretical measurement problem advocated within this paper corresponds to an element of indeterminacy in the meaning 3 of the above classification. Indeterminacy has no ontological status, at least not according to the description of nature given by quantum theory.
However, with respect to the description of nature by human beings there exists a fundamental element of indeterminacy being inevitable. It has its origin in the indispensability for human beings to describe physical objects as being located within a position space. All measurements, even measurements of internal variables are derived from measurements of a position. If the spin of a particle is measured within a Stern Gerlach experiment for example, the quantity being measured directly is the position which is performed by observing where there appears an absorbance. From this there is inferred the corresponding momentum which is directly correlated to the spin of a particle. 
The most fundamental objects in nature known so far are described by quantum theory implying states not describing trajectories in the kinematical sense of classical mechanics according to the uncertainty relation and have no sharp position and momentum at the same time. They are not even located to a small region, if they appear as free objects. Only within bound states within larger systems incorporating other objects there can be obtained a more sharply localization (within the borders defined by the uncertainty relation).
To enable the description as a classical system an object has to consist of such a large number of particles that it is impossible to know its state exactly. Space is a precondition for human perception. But only macroscopic objects fulfil the precondition to obey such a description. Therefore it is necessary to consider an interaction of an elementary particle with a macroscopic object to obtain knowledge about it. And this implies that there is only possible an indeterministic description. Due to the above argumentation this element of indeterminacy cannot be omitted. With respect to statistical mechanics it is at least thinkable that the exact state of every constituent of a complex system is measured and thus a deterministic description is not excluded in principle but by practical reasons. In quantum theory such a description is not even thinkable because to measure the states of all the particles a measurement instrument consists of one would need another macroscopic measurement instrument for which the same argumentation would be valid. An attitude being quite similar is described explicitly by Peter Mittelstaedt \cite{Mittelstaedt:1963}:

\begin{quote}
"`It belongs to these preconditions that there has to be used a measurement instrument of macroscopic dimension. Because during an interaction of this instrument with the object the information being available originally in compact form is distributed to the estimated $10^{20}$ degrees of freedom of the measurement instrument and becomes - at least to a certain extent - practically unavailable. This is because the calculational description of the complex equation systems which are given in the large state space of S+M [measured system and measurement instrument] extends the scope of the real possibilities completely."'

Peter Mittelstaedt, Philosophische Probleme der modernen Physik ("`Philosophical Problems in Modern Physics"'), 1963. 
\end{quote}
That the theory according to the attempt for a reconciliation of postulate 3 and 4 presented here is not deterministic in the sense that random appears not in an ontological but in an epistemological sense in a principle way, is logically completely independent of the interpretation of the uncertainty relation, especially the fact that it makes no sense to talk about an exact location and an exact momentum of a particle at the same time which extends the accuracy being determined by the uncertainty relation being a direct consequence of the postulates of quantum theory. Often the new sort of description of nature related to quantum theory is presented in such a way that in quantum theory one describes just probabilities and one gets something real, if one measures the position or the momentum. According to such an attitude the state has no reality by itself but just with respect to the observable quantities.
I think that by referring to the way the contradiction between postulate 3 and postulate 4 is reconciled it is much more plausible to interpret the state as a real entity. With respect to this interpretation the structure of the abstract Hilbert space is considered to be more fundamental and the representation within a space of positions or momenta is interpreted to be a derived one. Concerning this aspect one has to remember that the postulates of quantum theory according to Dirac and von Neumann \cite{Dirac:1958,Neumann:1932} do not presuppose a space of positions and momenta, not even in the sense of a representation space for the states which appear as wave functions in such a representation. Internal degrees of freedom for example have no relation to position space at all. 
Accordingly, postulate 3 describes a dynamical evolution of a state in Hilbert space which serves as a complete description
of a physical system without mandatorily referring to measurable quantities. This yields a strong justification of the opinion that states by themselves have a higher degree of reality than representations by observable quantities like position and momentum which are quantities being measurable directly. Concerning a general state being degenerated with respect to a certain observable quantity it makes no sense to assume that there exists a real value of this observable quantity and the state just describes the probability distribution. The state itself has to be considered as the reality. Therefore let us cite again Mittelstaedt \cite{Mittelstaedt:1963}:

\begin{quote}
"`The matter of fact that $|\varphi \rangle$ is the state of the system S, represents an objective property of S in the sense that the operator $P_\varphi=|\varphi \rangle \langle \varphi|$ has the measurement value 1 instead of 0. The coefficients 
$\langle A_i|\varphi \rangle=\varphi(A_i)$ are nothing else than a special representation of the state $|\varphi \rangle$ and indicate that the system S holds indeed the state $|\varphi \rangle$. $P_\varphi$ has thus to be considered as an objective description of the system S."'

Peter Mittelstaedt, Philosophische Probleme der modernen Physik ("`Philosophical Problems in Modern Physics"'), 1963. 
\end{quote}
This attitude holding that states have a higher ontological status than quantities being directly measurable gets important support by the mentioned fact that the phase of a quantum theoretical wave function in position space not even corresponds to probabilities of quantities being directly measurable. Nevertheless it is necessary for an explanation of certain effects of interference and therefore has to be considered as an ontological reality.

\section{The duality between particles and waves and the relation between the abstract Hilbert space of states and space-time}

In this section there shall be given a short comment on the duality between particles and waves and then it will be made the attempt to explore the consequences of the above considerations with respect to the ontological status of space within quantum theory.\\
I think that the duality between particles and waves has to be interpreted in such a way that particles being described by abstract states in Hilbert space which can be represented in position space are the fundamental entities of nature. Quantum fields should be interpreted as multi particle states. However, a particle is no localized object in the classical sense. It is described by a state obeying the uncertainty relation. This means that it would be more suitable to define a particle in the sense of a quantum which can be distinguished from other quanta in a discrete way .
And such a quantum can also be denoted as a wave in the sense that its state can be represented as a wave. The waves of quantum field theory have to be interpreted as composed objects from the fundamental particles which states are represented as waves, too. In the approximation of a delta function in position space the state itself gets the properties of a classical particle.\\
Concerning a real comprehension of the ontological meaning of the uncertainty relation and the corresponding non locality of quantum theory which becomes manifest especially with respect to the Einstein Podolsky Rosen paradox \cite{Einstein:1935} and the corresponding close relation between particles being separated by a large distance, there has to be considered an attitude
where non locality is essential and locality appears as a macroscopic phenomenon in nature. The general abstract formulation of quantum theory which does not refer to position or momentum space seems to enable such an attitude and Bohr's concept of individuality \cite{Bohr:1935af} seems to be a manifestation of this essential non locality. 
Therefore it is suggested that position space is just a kind of representation of relations between dynamical entities. With respect to this quantum theory corresponds to the philosophical implications about the nature of space-time within general relativity where the metric structure of space-time is described as a dynamical entity and coordinates lose their physical meaning according to the property of diffeomorphism invariance. Therefore it seems to be plausible that the structure of space time as a (3+1)-dimensional manifold is itself a kind of representation of an underlying structure of abstract states in Hilbert space. (This does not imply that time loses its special status.) Since quantum theory has to be assumed to be more general than classical physics, its concepts have to be interpreted as being more exact. The concept of particles moving in a space-time being an independent physical entity is a classical concept. But the structure of quantum theory in its general formulation refers to states in Hilbert space which may be represented approximatively in an 3-dimensional position space and the corresponding momentum space, respectively. The limits of validity of this representation is given by the uncertainty relation. The space of human imagination is a kind of representation of the physical relations within the real world. It is mandatory for human perception and it reflects and corresponds to the approximative structure of classical physics where the kinematics of objects can be described by trajectories through space with a sharp position and a sharp momentum at the same time although it is not completely isomorphic to the structure of the real world as it is shown by the uncertainty relation.

\section{Summary and discussion}

It has been made the attempt to understand the philosophical attitude of the Copenhagen interpretation of quantum theory which is assumed to be constitutive for a comprehension of the physical meaning of quantum theory, at least as a necessary condition. Concerning this it was decisive to emphasize the epistemological character of the arguments of Heisenberg and Bohr. Besides it has been advocated the opinion that the measurement problem which has been presented as the decisive conceptual problem of quantum theory can be solved by investigating the ontological status of the postulates referring to a changing of quantum states and their relation to each other. This has led to an interpretation of quantum theory which is related to the statistical character of statistical mechanics and arises from the necessity to use macroscopic measurement instruments which obey classical concepts to describe quantum physics. This necessity to use macroscopic objects corresponding to the approximative applicability of classical concepts has its origin in the constitution of human perception which is based on location within a position space.
According to this interpretation indeterminacy has no meaning in a strict ontological sense but in a principle epistemological sense which means that it is inevitable because of the relation and incommensurability of the description of nature according to quantum states obeying the uncertainty relation and the nature of human perception.
After all from the above argumentations there has been drawn the conclusion that the description of nature by objects within a position space reflects a certain structure which is derived from the more abstract and therefore more general description by quantum states of elementary objects whose real nature the concepts of a position and a momentum are not able to 
comprise in a suitable way.


\begin{thebibliography}{99}

\bibitem{Einstein:1935} A. Einstein, B. Podolsky and N. Rosen,
 ``Can Quantum-Mechanical Description of Physical Reality Be Considered Complete?,''
 Phys. Rev. 47, 777-780 (1935).

\bibitem{Weizsaecker:1971}
 C.~F.~von Weizsaecker, {\it Die Einheit der Natur},
 (Carl Hanser Verlag, Muenchen, 1971.)

\bibitem{Heisenberg:1958}
 W.~Heisenberg, {\it Physik und Philosophie},
 (S. Hirzel Verlag, Stuttgart, 2000.)
 {\it [original edition: Physics and Philosophy, Harper \& Brothers, New York, 1958.]}

\bibitem{Heisenberg:1969}
 W.~Heisenberg, {\it Der Teil und das Ganze},
 (Piper Verlag, Muenchen, 1969.)

\bibitem{Heisenberg:1927}
 W.~Heisenberg,
 ``Ueber den anschaulichen Inhalt der quantenmechanischen Kinematik und Dynamik,''
 Zeitschrift fuer Physik, Volume 43, Numbers 3-4, pages 172-198 (1927).

\bibitem{Weizsaecker:1999}
 C.~F.~von~Weizsaecker, {\it Grosse Physiker},
 (Deutscher Taschenbuchverlag, Muenchen 1999.)

\bibitem{Bell:1964fg}
 J.~S.~Bell,
 ``On The Problem Of Hidden Variables In Quantum Mechanics,''
 Rev.\ Mod.\ Phys.\  {\bf 38} (1966) 447.

\bibitem{Kant:1781}
 I.~Kant, {\it Kritik der reinen Vernunft}, 1781.

\bibitem{Kant:1783}
 I.~Kant, {\it Prolegomena zu einer jeden kuenftigen Metaphysik, die als Wissenschaft wird auftreten koennen}, 1783.

\bibitem{Lorenz:1970}
 K.~Lorenz,
 {\it Die Rueckseite des Spiegels, Versuch einer Naturgeschichte menschlichen Erkennens}, Muenchen 1973.

\bibitem{Vollmer:1975}
 G.~Vollmer,
 {\it Evolutionäre Erkenntnistheorie}, Stuttgart 1975.

\bibitem{Dirac:1958}
  P.~A.~M.~Dirac,
  {\it The principles of quantum mechanics}, Oxford University Press, Oxford 1958.

\bibitem{Neumann:1932}
  J.~von~Neumann,
  {\it Mathematische Grundlagen der Quantenmechanik}, Berlin 1932.

\bibitem{Weizsaecker:1985}
 C.~F.~von~Weizsaecker, {\it Aufbau der Physik}, Carl Hanser Verlag, Muenchen, 1985.

\bibitem{Weizsaecker:1992}
 C.~F.~von~Weizsaecker, {\it Zeit und Wissen}, Carl Hanser Verlag, Muenchen, 1992.

\bibitem{Rovelli:1995fv}
 C.~Rovelli,
 ``Relational Quantum Mechanics,''
 arXiv:quant-ph/9609002.

\bibitem{Rovelli:2004}
 C.~Rovelli, {\it Quantum Gravity}, (Cambridge University Press, 2004.)

\bibitem{Smerlak:2006gi}
 M.~Smerlak and C.~Rovelli,
 ``Relational EPR,''
 Found.\ Phys.\  {\bf 37} (2007) 427
 [arXiv:quant-ph/0604064].

\bibitem{Schlosshauer:2003zy}
 M.~Schlosshauer,
 ``Decoherence, the Measurement Problem, and Interpretations of Quantum
 Mechanics,''
 Rev.\ Mod.\ Phys.\  {\bf 76} (2004) 1267
 [arXiv:quant-ph/0312059].

\bibitem{Schlosshauer:2008}
 M.~Schlosshauer, K.~Camilleri,
 ``The quantum-to-classical transition: Bohr's doctrine of classical concepts, 
 emergent classicality, and decoherence,''
 [arXiv:0804.1609]. 

\bibitem{Mittelstaedt:1963}
 P.~Mittelstaedt, {\it Philosophische Probleme der modernen Physik}, (Bibliographisches Institut, Mannheim, 1963.)

\bibitem{Bohr:1935af}
 N.~Bohr,
 ``Can Quantum-Mechanical Description of Physical Reality be Considered Complete?,''
 Phys.\ Rev.\  {\bf 48} (1935) 696.

\end{thebibliography}
\end{document}